\documentclass[seceq]{ptptex}

\usepackage{graphicx}
\title{Mesons in Nuclei: $\eta$ and $\omega$ mesons}
\author{Susan \textsc{Schadmand}}
\inst{IKP \& JCHP, Foschungszentrum J\"ulich, Germany}

\abst{%
Data on the photoproduction of $\omega$ mesons on nuclei have been re-analyzed in search for
in-medium modifications. The data were taken with the Crystal Barrel(CB)/TAPS detector system at
the ELSA accelerator facility in Bonn.  The extracted $\omega$ line shape was found to be sensitive
to the background subtraction.

In experiments at the tagged photon facility of the Mainz MAMI accelerator photoproduction of
mesons from light nuclear targets (deuteron and $^3\mbox{He}$) has been studied. The experiments
used the combined Crystal Ball/TAPS setup in Mainz. Measurements of $\eta$- and
$\pi^0$-photoproduction off a liquid $^3\mbox{He}$-target have been used for the search for the
formation of $\eta$-mesic $^3\mbox{He}$.

The installation of the WASA detector at COSY opened a unique possibility to search for the
$^4{\mbox{He}}-\eta$ bound state with  high statistics and high acceptance. We are conducting a
search via an exclusive measurement of the excitation function for the $dd \rightarrow
{^3\mbox{He}}\, p\, \pi^-$ reaction.
}

\PTPindex{201}  %Input the subject index(es) of your paper,
\begin{document}

\maketitle

\section{The $\omega$ Meson in Medium}

A possible modification of hadron properties in a strongly interacting medium is a much debated
issue. In-medium changes of hadron properties have been identified as one of the key problems in
understanding the non-perturbative sector of QCD. Fundamental symmetries in QCD provide guidance in
dealing with strong interaction phenomena in the non-perturbative domain. Furthermore, QCD sum
rules have been applied to connect the quark-gluon sector to hadronic descriptions and QCD-inspired
hadronic models have been developed to calculate the in-medium self-energies of hadrons and the
spectral functions. Mass shifts and/or in-medium broadening as well as more complex structures in
the spectral function due to the coupling of vector mesons to nucleon resonances have been
predicted. A recent overview is given in~Ref. \citen{Rapp:2009yu}. The studies have motivated several
experimental attempts to confirm or refute theoretical predictions.

Heavy-ion collisions and reactions with photons and protons have been used to extract experimental
information on in-medium properties of hadrons. The experiments have focused on the light vector
mesons $\rho, \omega$ and $\phi$ where decay lengths are comparable to nuclear dimensions in a
nuclear reaction. However, in order to ensure a reasonable decay probability in the strongly
interacting medium, cuts on the recoil momentum are required for the longer lived $\omega$ and
$\phi$ mesons.
A full consensus has not yet been reached among the different experiments. A detailed account of
the current status of the field is given in comprehensive reviews
\cite{Hayano:2008vn,Leupold:2009kz}.
An in-medium broadening of the vector mesons is reported by almost all experiments and the majority
of experiments does not find evidence for a mass shift. Apart from Ref. \citen{Trnka:2005ey} only one
other experiment at KEK \cite{Naruki:2005kd} reports a possible drop of the $\rho$ and $\omega$
mass by 9 $\%$ at normal nuclear matter density. Studying $\omega$ meson production in
ultra-relativistic heavy-ion collisions, the NA60 collaboration observes a suppression of the meson
yield for $\omega$ momenta below 1~GeV/c which is even more pronounced for more central collisions
\cite{Damjanovic:2006bd}. This is interpreted as evidence for in-medium modifications of slow
$\omega$ mesons but it cannot be concluded whether this is due to a mass shift, a broadening, or
both.

The search for mass shifts has turned out to be more complicated than initially thought for those
cases where a strong broadening of the meson is observed as for the $\omega$
\cite{Kotulla:2008xy} and $\phi$ meson \cite{Ishikawa:2004id}. In the $\omega \rightarrow \pi^0
\gamma$ decay mode the increase in the total width of $\omega$ drastically lowers the branching
ratio for in-medium decays into this  channel and thereby reduces the sensitivity of the observed
$\omega$ signal to in-medium modifications.
Data taken with the CB/TAPS detector at ELSA, Bonn on the photoproduction of $\omega$ mesons on
$Nb$ and $LH_2$ have been re-analyzed. The first results from an analysis of the same data were
published by D. Trnka et al. \cite{Trnka:2005ey}, claiming a mass shift of the $\omega$ meson by
-14$\%$ at normal nuclear matter density. This information was extracted from a comparison of the
$\omega$ signals on $Nb$ and $LH_2$, reconstructed in the $\pi^0 \gamma$ channel. As pointed out in
the literature \cite{Kaskulov:2006zc} the deduced line shapes are very sensitive to the
background subtraction. While in the initial work the background was determined by fitting the
$\pi^0 \gamma $ invariant mass spectrum a much more refined background determination is used in the
current analysis.
The re-analysis is described in detail in~Ref. \citen{Nanova:2010sy} where it is also stated that the
earlier claim of an in-medium lowering of the $\omega$ mass is not confirmed. The strong broadening
of the $\omega$ meson in the nuclear medium, due to inelastic processes as determined in the
transparency ratio measurements, suppresses contributions to the observed $\omega$ signal from the
interior of the nucleus. The branching ratio for in-medium decays into the channel of interest is
drastically reduced. Thereby, the sensitivity is shifted to the nuclear surface, making the line
shape analysis less sensitive to a direct observation of in-medium modifications. Data with much
higher statistics will be needed to gain further insight. A corresponding experiment has been
performed at the MAMI~C electron accelerator using the Crystal Ball/TAPS detector setup. The
analysis is ongoing.

\section{Search for Light $\eta$ Mesic Nuclei}

\subsection{Photoproduction of Mesons from Light Nuclei}

The study of the interaction of mesons with nucleons and nuclei has significantly contributed to
the understanding of the strong force. In the case of long-lived mesons like charged pions or
kaons, secondary beams can be prepared. However, the interaction of short-lived mesons with nuclei
is only accessible in indirect ways when the mesons are first produced in the nucleus from the
interaction of an incident beam and then subsequently undergo final state interactions in the same
nucleus. The pion-nucleon interaction at small momenta is weak, so that pion-nucleus states,
bound due to the strong force, cannot exist. The strong interaction is not only weak but even
repulsive at small momenta because of the s-wave term dominance.
However, the $\eta$-N interaction at small momenta is dominated by the s-wave nucleon resonance
S$_{11}$(1535), which couples strongly to the N$\eta$-channel~\cite{Krusche:1997jj}. The possible
formation of $\eta$ - nucleus bound states, $\eta$-mesic nuclei, has been intensively discussed in
the literature. The first suggestion for A$>$10 nuclei goes back to Liu and Haider~\cite{Liu_86}.
Even lighter quasi-bound $\eta$-nucleus systems have been sought for in experiments investigating
the threshold behavior of hadron induced $\eta$-production reactions. Quasi-bound states in the
vicinity of the production threshold will give rise to an enhancement of the cross section relative
to phase space behavior. Results were reported for
$pp\to\,pp\eta$~\cite{Calen:1996mn,Smyrski:1999jc,Moskal:2003gt},
$np\to\,d\eta$~\cite{Plouin:1990uj,Calen_98}, $pd\to\,^3\mbox{He}\eta$~\cite{Mayer:1995nu},
$dp\to\,^3\mbox{He}\eta$~\cite{Smyrski:2007nu,Mersmann:2007gw},
$\vec{d}d\to\eta\,^4\mbox{He}$~\cite{Willis:1997ix}, and $pd\to\,pd\eta$~\cite{Hibou:2000vp}.
Threshold effects have been found in most reactions, in fact, the measurement of
$dp\to\,^3\mbox{He}\eta$ by Mersmann et al.~\cite{Mersmann:2007gw} shows an extremely sharp rise
of the production cross section at threshold. If these effects are related to (quasi-)bound states,
they should show up independent of the initial state of the reaction. Threshold photoproduction of
$\eta$-mesons from light nuclei (deuteron, helium isotopes) was also
investigated~\cite{Hejny:2002vt,Pfeiffer:2003zd} and again threshold enhancements were observed.

The most promising signal so far was found in the $^3\mbox{He}(\gamma,\eta)^3\mbox{He}$
reaction~\cite{Pfeiffer:2003zd}, although the statistical significance was weak. Photoproduction
off $^3\mbox{He}$ has previously been investigated with the TAPS detector at the MAMI accelerator
in Mainz, Germany, and some indication for the existence of an $\eta$-mesic state has been found.
The cross section of coherent $\eta$-production showed a strong threshold enhancement. The
excitation function of $p\pi^0$ back-to-back pairs in the photon-helium cm-system, which may arise
from the $S_{11}\to\,N\pi$ decay after re-capture of the quasibound $\eta$ by a proton, showed a
peak around the $\eta$-production threshold. However, both effects suffered from poor statistical
quality of the data. The experiment has now been repeated with the much larger acceptance of the
new 4 detector setup. Preliminary results are reported in~Ref. \citen{Krusche:Messina09}. The
excitation function for coherent $\eta$-photoproduction is very similar to the results from the
hadron induced reaction~\cite{Mersmann:2007gw} where the cross section rises abruptly at the
coherent production threshold. Within the first 4~MeV above threshold it reaches roughly 50\% of
its maximum value, which is strong evidence for a resonance-like threshold enhancement. Data with
better resolution of the incident photon energy are still under analysis. The ongoing analysis is
revealing that nucleon resonances produce opening angle dependent structures in excitation
functions. Therefore, the method of subtracting the excitation functions for different opening
angles can produce artificial structures making it very difficult to isolate the small signature
from an $\eta$-mesic state in such a complicated landscape.

\subsection{Search for a $\eta-^4\mbox{He}$ State with WASA-at-COSY}

The installation of the WASA detector~\cite{Bargholtz:2008ze} at the COSY accelerator in
J\"ulich, Germany, opened a unique possibility to search for the $^4\mbox{He}\,-\eta$ bound state
with high statistics and high acceptance. A first exclusive measurement of the excitation function
for the $dd\rightarrow\,^3\mbox{He}\,p\,\pi^-$ reaction. Here, ramping of the beam momentum was
used to continuously vary the beam momentum around the threshold of the
$d\,d\to\,^4\mbox{He}\,\eta$ reaction. The $^4\mbox{He}\,-\,\eta$ bound state should manifests
itself as a resonance-like structure below the threshold for the $dd\to\,^4\mbox{He}\,\eta$
reaction. If a peak below the $^4\mbox{He}\,\eta$ threshold is found then the profile of the
excitation curve will allow to determine the binding energy and the width of the
$^4\mbox{He}\,-\,\eta$ bound state. However, if only an enhancement around the threshold is found,
then it will enable to establish the relation between width and binding
energy~\cite{GarciaRecio:2002cu}. In the absence of a visible structure, the upper limit for the
cross section of the production of the $\eta$-helium nucleus will be a few $nb$. The search for a
signature of the $\eta$-mesic state takes advantage of the fact that the distribution of the
opening angle of the nucleon-pion pair for the background is much broader than the one expected
from the decay of the bound state. This is because the opening angle of the outgoing nucleon-pion
pair originating from the decay of the N$^*(1535)$ resonance is equal to 180$^{\circ}$ in the N$^*$
reference frame and it is smeared only by about 30$^{\circ}$ in the reaction center-of-mass frame
due to the Fermi motion of the nucleons inside the $\mbox{He}$ nucleus. This allows for comparing
excitation functions corresponding to the "signal-rich" and "signal-poor" regions.

\begin{figure}
  \centerline{\includegraphics[width=\textwidth]{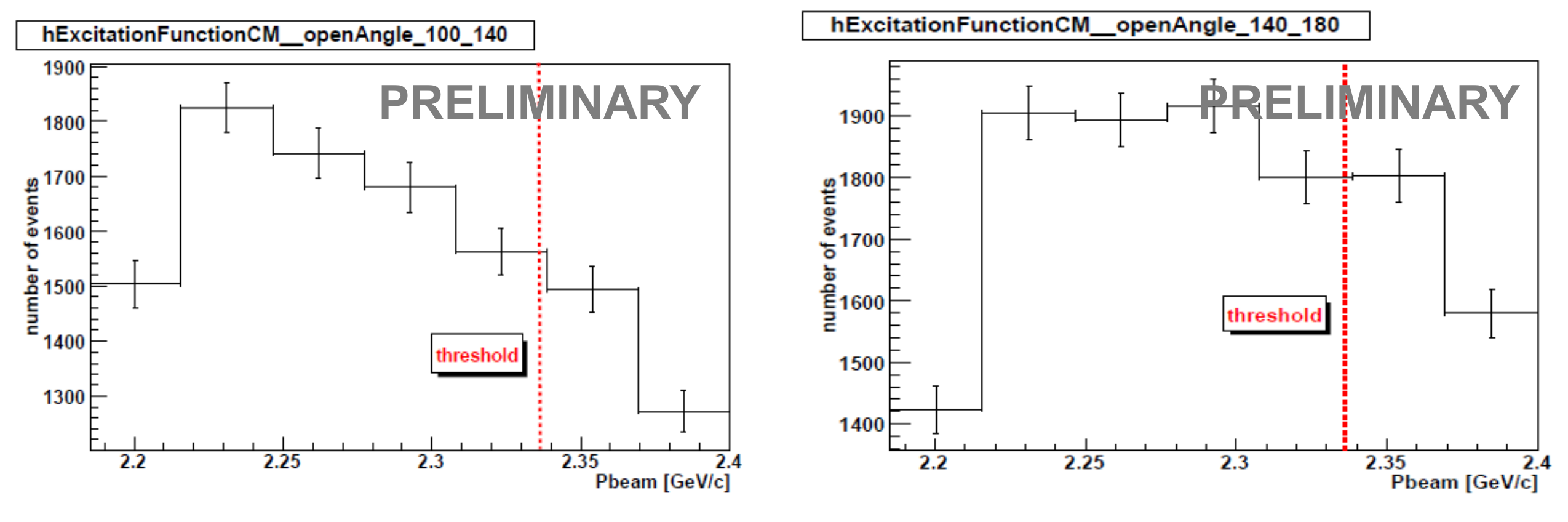}}
  \caption{Preliminary comparison of excitation functions around the $\eta$ threshold in the
  reaction $dd\rightarrow\,^3\mbox{He}\,p\,\pi^-$. The data have not yet been corrected for possible
  luminosity variations during the COSY cycle.
  Right panel: Excitation function for the $p\pi$ opening angle in the center-of-mass frame
  for opening angles between 140 and 180 degrees. Note that the zero of the ordinate is suppressed.
  Left: for opening angles between 100 and 140 degrees.
    }\label{fig:Krzemien}
\end{figure}

The pilot experiment, conducted in June~2008, used a deuteron pellet target and a deuteron beam
with a ramped momentum corresponding to a variation of the excess energy for the
$^4\mbox{He}\,-\,\eta$ system from -51.4~MeV to 22~MeV. At present the data are evaluated and some
preliminary results are reported~\cite{Krzemien:2009wm}.
In figure~\ref{fig:Krzemien}, a preliminary comparison of the excitation function for two $p\pi$
opening angular ranges are shown.
Barring the lack of statistics from this pilot experiment, a difference in the shape of the spectra
can be discerned. It will have to be seen whether this indication can be confirmed in the planned
continuation of the experiment. Furthermore, particular attention will be paid to possible
background from single pion production as referred to in the previous subsection. The experiment
will be continued in 2010~\cite{Adam:2004ch} with a two week run scheduled with WASA-at-COSY.

\section*{Acknowledgements}
Discussions during the YIPQS international workshop on "New Frontiers in QCD 2010", were very
useful for this work.

Parts of this publication have been supported by the European Commission under the 7th Framework
Programme through the 'Research Infrastructures' action of the 'Capacities' Programme. Call:
FP7-INFRASTRUCTURES-2008-1, Grant Agreement N. 227431.

%--------------------------------------------


\begin{thebibliography}{10}

\bibitem{Rapp:2009yu}
R.~Rapp, J.~Wambach, H.~van Hees, hep-ph/0901.3289.

\bibitem{Hayano:2008vn}
R.~S. Hayano, T.~Hatsuda, nucl-ex/0812.1702.

\bibitem{Leupold:2009kz}
S.~Leupold, V.~Metag, U.~Mosel, Int. J. Mod. Phys. E19 (2010) 147--224.

\bibitem{Trnka:2005ey}
D.~Trnka, et~al., Phys. Rev. Lett. 94 (2005) 192303.

\bibitem{Naruki:2005kd}
M.~Naruki, et~al., Phys. Rev. Lett. 96 (2006) 092301.

\bibitem{Damjanovic:2006bd}
S.~Damjanovic, Eur. Phys. J. C49 (2007) 235--241.

\bibitem{Kotulla:2008xy}
M.~Kotulla, et~al., Phys. Rev. Lett. 100 (2008) 192302.

\bibitem{Ishikawa:2004id}
T.~Ishikawa, et~al., Phys. Lett. B608 (2005) 215--222.

\bibitem{Kaskulov:2006zc}
M.~Kaskulov, E.~Hernandez, E.~Oset, Eur. Phys. J. A31 (2007) 245--254.

\bibitem{Nanova:2010sy}
M.~Nanova, et~al., nucl-ex/1005.5694.

\bibitem{Krusche:1997jj}
B.~Krusche, N.~C. Mukhopadhyay, J.~F. Zhang, M.~Benmerrouche, Phys. Lett. B397
  (1997) 171--176.

\bibitem{Liu_86}
L.C.Liu, Q.Haider, Phys. Rec. C34 (1986) 1845.

\bibitem{Calen:1996mn}
H.~Calen, et~al., Phys. Lett. B366 (1996) 39--43.

\bibitem{Smyrski:1999jc}
J.~Smyrski, et~al., Phys. Lett. B474 (2000) 182--187.

\bibitem{Moskal:2003gt}
P.~Moskal, et~al., Phys. Rev. C69 (2004) 025203.

\bibitem{Plouin:1990uj}
F.~Plouin, P.~Fleury, C.~Wilkin, Phys. Rev. Lett. 65 (1990) 690--693.

\bibitem{Calen_98}
H.~Cal\'en, et~al., Phys. Rev. Lett. 80 (1998) 2069.

\bibitem{Mayer:1995nu}
B.~Mayer, et~al., Phys. Rev. C53 (1996) 2068--2074.

\bibitem{Smyrski:2007nu}
J.~Smyrski, et~al., Phys. Lett. B649 (2007) 258--262.

\bibitem{Mersmann:2007gw}
T.~Mersmann, et~al., Phys. Rev. Lett. 98 (2007) 242301.

\bibitem{Willis:1997ix}
N.~Willis, et~al., Phys. Lett. B406 (1997) 14--19.

\bibitem{Hibou:2000vp}
F.~Hibou, et~al., Eur. Phys. J. A7 (2000) 537--541.

\bibitem{Hejny:2002vt}
V.~Hejny, et~al., Eur. Phys. J. A13 (2002) 493--499.

\bibitem{Pfeiffer:2003zd}
M.~Pfeiffer, et~al., Phys. Rev. Lett. 92 (2004) 252001.

\bibitem{Krusche:Messina09}
B.~Krusche, to be published in Int. J. Mod. Phys. E.

\bibitem{Bargholtz:2008ze}
C.~Bargholtz, et~al., Nucl. Instrum. Meth. A594 (2008) 339--350.

\bibitem{GarciaRecio:2002cu}
C.~Garcia-Recio, J.~Nieves, T.~Inoue, E.~Oset, Phys. Lett. B550 (2002) 47--54.

\bibitem{Krzemien:2009wm}
W.~Krzemien, P.~Moskal, J.~Smyrski, Acta Phys. Polon. Supp. 2 (2009) 141--148.

\bibitem{Adam:2004ch}
H.~H. Adam, et~al., {Proposal for the Wide Angle Shower Apparatus (WASA) at
  COSY-Juelich - 'WASA at COSY'}, nucl-ex/0411038.

\end{thebibliography}
\end{document}